%% file: main.tex
\documentclass[final,12pt]{elsivier/elsarticle}

\usepackage{amssymb}
\usepackage{xspace}
\usepackage{url}
\usepackage{bigstrut}
\usepackage{graphicx}
\usepackage{breakurl} 
\usepackage{hyperref}
\hypersetup{colorlinks,allcolors=black}

\renewcommand{\address}[1]{\texttt{#1}\xspace}
\newcommand{\qMig}{\textsc{qMig}\xspace}
\newcommand{\FailSafe}{\textsc{FailSafe}\xspace}
\newcommand{\point}[1]{\smallskip\par\noindent\textbf{#1:}}
\newcommand{\code}[1]{\ensuremath{\mathtt{{#1}}}}

\journal{Blockchain: Research and Applications}

\begin{document}

\begin{frontmatter}



\title{From Social Engineering to Quantum Threats: Safeguarding User Wallets with FailSafe}


\author[1]{Gennady ``Ari'' Medvinsky}

\affiliation[1]{organization={Eleos Labs}},

\author[1,2]{Benjamin Livshits}

\affiliation[2]{organization={Imperial College London}}

\begin{abstract}
While cryptocurrencies have been rapidly gaining adoption, secure wallet interactions are still elusive for many users, which frequently leads to loss of funds. Here we propose an approach to securing interactions with cryptocurrency wallets for end-users. The approach called \FailSafe consists of several defence-in-depth measures that can be applied near-term as well as a tool called \qMig for aiding eventual quantum migration.
\end{abstract}



\end{frontmatter}



\input{sections/01_introduction.tex}
\input{sections/02_threats.tex}

\input{sections/03_architecture.tex}
\input{sections/04_related.tex}
\input{sections/05_conclusions.tex}


\bibliographystyle{elsarticle-num} 

\bibliography{biblio}





\end{document}

%% file: sections/01_introduction.tex
\section{Introduction}
\label{sec:intro}

In Web2, most users have a certain degree of familiarity with threats to their online accounts, be those of financial nature or pertaining to other services like social media. Frequently, the attacker focuses on user credentials and account takeover. Password compromises can be achieved through a dictionary attack, taking advantage of a poorly chosen secret, or through many forms of phishing and social engineering attacks, enticing the user to share their credentials. With the password  compromised, the  attacker can then extract whatever value is associated with the account, or use that account to perform some form of escalation to compromise further accounts.  According to recent surveys~\cite{editorfebruary19_2020_users},  the following best online security practices remain a challenge for both end-users and IT professionals alike.

Usability issues around security have some parallels in Web3.  For instance, phishing attacks can vary from tricking the user to disclose their private key, to obtaining a signature that grants permission for potentially unlimited fund transfer.  Flawed but popular open-source software can generate vanity blockchain addresses (i.e, Profanity~\cite{group_2022_exploiting}), but in the process makes it trivial for the attacker to compute the corresponding private key. 

Most products in this space are built with relatively standalone threat models in mind.  For example, hardware wallet solutions tend to mainly focus on having an air gap with less trusted software components. Threat intelligence products like Chainalysis provide an insight into the counter-party risk.  Users are left to devise their own threat mitigation strategies through some combination of these products. This approach not only puts an undue burden on the user, but may leave users exposed to web3 threats in unforeseen ways.

\begin{figure}[bt]
\centering
\includegraphics[width=0.75\textwidth]{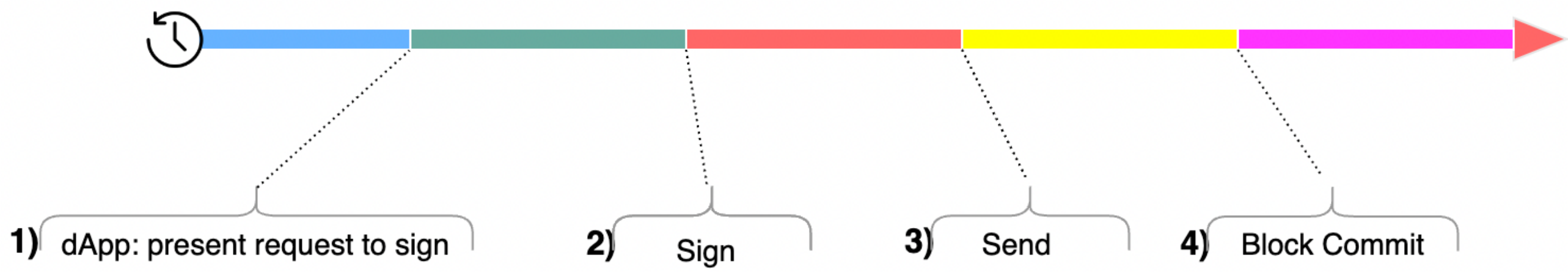}
\caption{Timeline of a Transaction.}
\label{fig:tx-timeline}
\end{figure}

\subsection{Defence-in-Depth}
\FailSafe is an anti-theft Web3 wallet companion system that is focused on protecting the end to end web3 transaction journey.  \FailSafe is built using the defence-in-depth principle: it offers a multilayered set of security mechanisms, with built-in redundancy, designed to minimise the loss of user assets even under the worst-case circumstances (disclosure of the user’s private key, or a compromised insider within a trusted system). 

\FailSafe takes every opportunity to protect the user’s assets across the lifecycle of a transaction: from initial user engagement phase with the dApp, to the point it is committed on chain.  At the outset, on enrollment, \FailSafe helps the user to reduce risk by moving the majority of assets to the user’s cold wallet address that does not partake in regular web3 transactions; this is not unlike what custody solutions do; however, up til now, this practice has been unavailable to retail users.  

According to a recent study~\cite{WangFWZ0Y22} of ERC-20 token usage patterns,~60\% of all users grant unlimited transfer approvals to dApps,~22\% of  which are considered to be at high risk of their approved tokens being stolen. By moving the majority of assets to the user’s cold wallet, these assets are no longer exposed to the above risk.

\begin{figure}[t]
\centering
\begin{tabular}{|ll|}
\hline
\bigstrut
     FBR &  counter-party risk \\
     FIS &  mempool operations \\
     \qMig & quantum migration \\
\hline
\end{tabular}
\caption{Layers of protection in \FailSafe.}
\label{fig:protections}
\end{figure}

\FailSafe, automatically maintains the user desired balance ratio between the hot and cold wallet addresses, preserving the de-risked security posture over time.  Once the user engages with a dApp, the \FailSafe Blockchain Reconnaissance~(FBR) service is used to obtain the risk score for the counter-party’s web3 address.  

If \FailSafe software is in the code path, fraudulent transactions are outright blocked. Otherwise, the next layer of protection is the \FailSafe Interceptor Service~(FIS), which monitors pending transactions submitted to the blockchain’s memory pool.  If the transaction counter-party has a high risk score (based on a call to FBR) , FIS is capable of submitting another transaction that is executed ahead of the attacker’s, moving the funds at risk into the user’s cold storage address before the attacker's transaction is executed.

\subsection{Forward Security}
\FailSafe defence-in-depth approach is forward-looking~---~it lays the groundwork for safeguarding the user’s crypto against newly-emerging threats.   

Advances in quantum computing hardware have made significant strides, propelled by the nation-state quantum computing race with a number of different R\&D centres, reaching significant computing benchmarks and milestones (see: Google’s Quantum Supremacy~\cite{48651_google} and IBM Quantum System One~\cite{a2015_ibm}).

When viewed from the lens of cryptography, it presents a unique problem. While Shor’s algorithm published in~1994~\cite{shor_polynomial-time_1997}, could theoretically break certain algorithms used for digital signatures (i.e., ECDSA), it requires a sufficiently powerful quantum computer to do it.  With recent advances, the time window to reach this milestone has been shrinking (see Global Risk Institute’s 2022 report~\cite{a2022_timeline}). 

The situation is especially dire for the Ethereum ecosystem (this includes EVM compatible networks, like Polygon, Binance Smart Chain, Avalanche, and many others).  The current version of Ethereum lacks cryptographic agility. Externally-owned addresses (user wallets) use ECDSA with no other option built in (see the quantum threats section for a more in-depth discussion).  
Furthermore, by design, externally owned addresses are commonly re-used, giving an attacker with future quantum hardware a longer time window to derive the private key via the earlier record of transaction signatures.

Once wallet signatures are no longer cryptographically trustworthy, the inability to establish rightful custody over web3 assets will pose a barrier to bridging  assets  to a quantum-safe network (e.g., QRL~\cite{qrl}) or future versions of quantum-safe, EVM-compatible blockchains.  

As part of the \FailSafe project, the Quantum Migration Tool~(\qMig) was developed to future-proof against this outcome.  Prior to the quantum inflection point, \qMig enables users to construct and record a future intent to transfer tokens, in case the quantum inflection point occurs and ECDSA signature by itself can not be trusted. The security of this intent is rooted in cryptography that is not susceptible to quantum attacks. The integration of \FailSafe with \qMig, records the necessary proofs automatically, requiring no additional effort by the end user, as detailed in Section~\ref{sec:qmig}.

%% file: sections/02_threats.tex
\begin{figure}[tbp]
\centering
\includegraphics[width=0.85\textwidth]{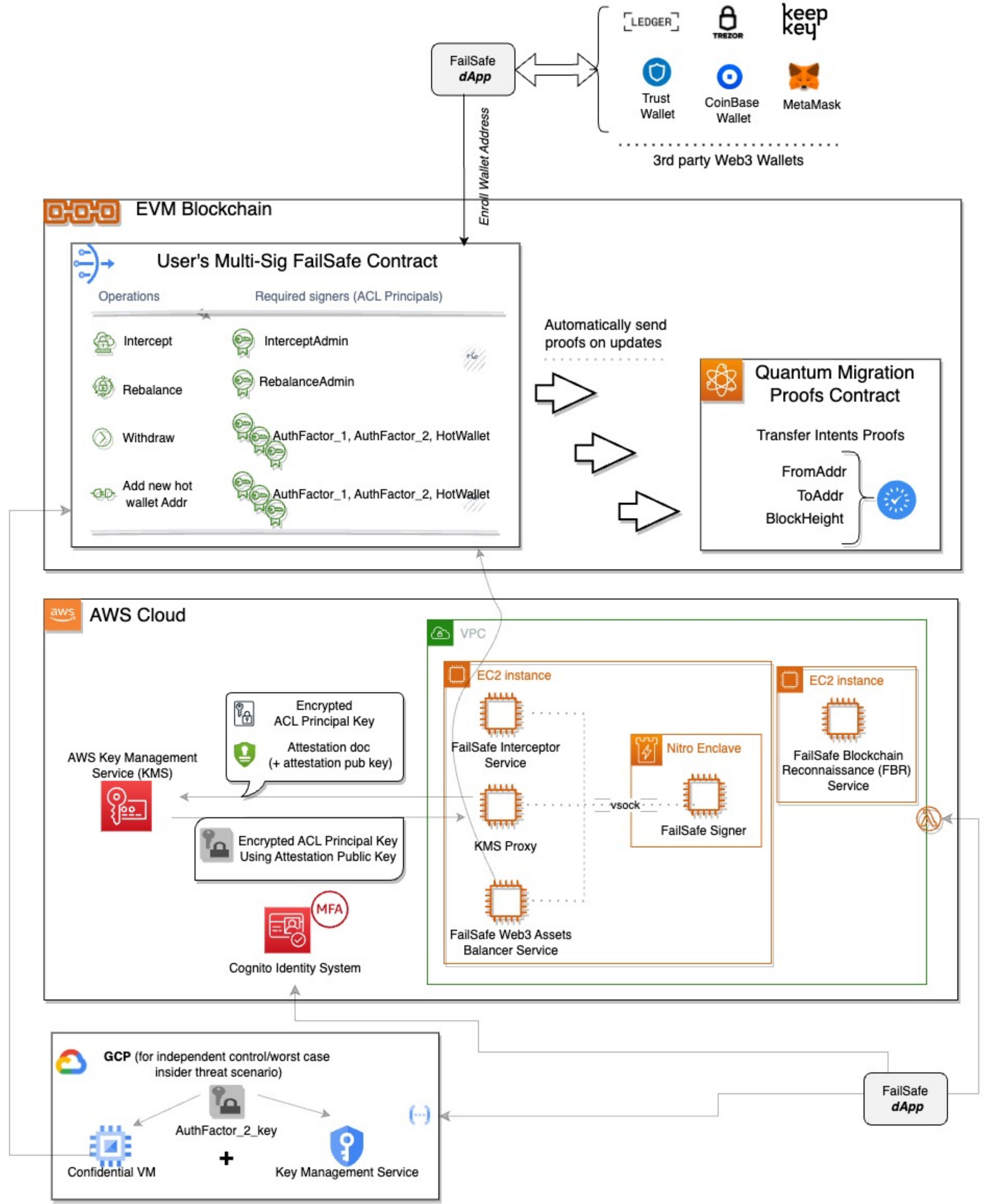}
\caption{\FailSafe architecture.}
\label{fig:arch}
\end{figure}

\section{Web3 Threats to Your Crypto}
\label{sec:threats}

The private key that corresponds to the user’s wallet address controls the transfer of value on the public ledger, be it in the form of tokens or native cryptocurrency. To capture this key, a potential attacker has a range of possible options; the reader is referred to Mirza~et.~al. for more details~\cite{forensics-wallets}:

\begin{itemize}
    \item \textbf{Theft of private keys}: with the knowledge of the private key, the attacker can send a transaction for every token and native currency associated with the address, transferring the assets to the attacker's own address. Any staked tokens in third party systems can be withdrawn and transferred to the attacker’s address. There are numerous examples of this in the wild: fraudsters often pose as customer support convincing users to install a fake wallet software that captures and shares the user’s passphrase with the attacker. Some of the recent bridge hacks have the same culprit as well~\cite{bridge-hacks}. 
    \item \textbf{Obtaining user’s authorization}:  through sociael engineering and confusing Web3 wallet user experiences, the attacker convinces the user to sign a transaction that can be crafted to transfer funds out of the wallet~\cite{WangFWZ0Y22}.
    \item \textbf{Compromise of third-party smart contracts}:  Exploit smart contract vulnerabilities and then drain user assets that temporarily reside under the contract address ownership (there are numerous examples of bridge hacks that fall into this category). 
\end{itemize}

\subsection{Designing with Operator Error in Mind}
From recent trends in Web3 attacks, it is clear that the human factor plays a central role.  Users might be lured into violating one or more best security practices without knowing – in the case of the Profanity bug hack, vanity Web3 addresses were generated that made it possible for attackers to derive the private key. Once the system is configured into a secure state, over time it is likely that the security posture will decay, if it requires regular end-user effort to upkeep.

In a recent~\$8M exploit, users were lured into installing an unofficial update of a popular web3 wallet.  It is suspected that the fake wallet update involved users re-entering the seed phrase (giving the attacker full access to the victim’s crypto assets). The \FailSafe threat model is designed with these seemingly game over scenarios in mind.  In the later part of this section, we introduce how the defense-in-depth principle is applied throughout the life-cycle of a transaction, and how the application of \FailSafe multi-layered defenses minimises losses from the type of incidents noted above.

\subsection{Defence-in-depth and the Lifecycle of a Transaction}
\FailSafe is built on the defence-in-depth principle: a multilayered set of security mechanisms, with built in redundancy, designed to minimise loss of user assets even in the worst case circumstances (e.g., user is tricked into giving away the wallet’s passphrase). A summary is shown in Figure~\ref{fig:protections}.

To better understand how this works, let's take a closer look at the life-cycle of a transaction: from initial user engagement phase with the dApp, to the point it becomes part of a permanent record on a public ledger (as illustrated in Figure~\ref{fig:tx-timeline}).

Each phase below presents both an opportunity for the attacker, as well as a chance to employ a countermeasure.

\point{Defence 1: de-risk Web3 Asset Positions}
Before engaging with the user, the attacker has an opportunity to learn a great deal from the public ledger, fine-tuning targets of interest,  based on type and value of owned assets.  From the public ledger, the attacker’s bot can compile a list of addresses and corresponding owned tokens on selected EVM blockchains, customising the attack as needed. 

On the flipside,  during this phase, the user has a chance to de-risk and remove the majority of owned assets entirely beyond the attacker’s reach.  By enrolling in the\FailSafe automated cold storage feature, the vast majority of assets are re-balanced, to be owned by the user’s wallet address that does not partake in regular web3 transactions.  

Just as importantly,\FailSafe is designed to maintain this security posture over time. With little to no imposition on the user,\FailSafe automatically maintains the asset balance ratio between the hot and cold wallet, subject to the user's high level instructions.  Access to cold storage is safeguarded via a multi-signature contract, the corresponding private keys are protected under a unique orchestration of Amazon’s Nitro Enclaves and Google’s Confidential Compute with cloud hardware.  
Just as importantly,\FailSafe is designed to maintain this security posture over time. With little to no imposition on the user,\FailSafe automatically maintains the asset balance ratio between the hot and cold wallet, subject to the user's high level instructions.  Access to cold storage is safeguarded via a multi-signature contract, the corresponding private keys are protected under a unique orchestration of Amazon's Nitro Enclaves and Google’s Confidential Compute with cloud hardware security modules~(HSM); it is designed to withstand insider threat/compromise (Figure~\ref{fig:arch} illustrates the overall architecture and described in more detail in the later section below). 

\point{Defence 2: \FailSafe Blockchain Reconnaissance: 
First contact with the attacker}.  As noted earlier, the attacker’s goal at this point is  either to directly learn the user's private key, or convince the user to sign a transaction of the attacker's choice.  A myriad of tried and tested social engineering attacks are available, and just as in the Web2 world, even if rejected by 99

At this stage, a \FailSafe user is protected by several countermeasures.  When the user encounters the attacker’s dApp, if the user is using a client that is directly integrated with \FailSafe Blockchain Reconnaissance~(FBR) service (e.g., like a \FailSafe Chrome extension or a proxy RPC URL), the attacker's request is likely to be rejected outright.  The FBR maintains an up-to-date database of black listed addresses; this includes sanctioned addresses, fraudulent/rug-pull contracts. Risk profiles are also constructed based on historical as well real time transaction driven behaviour anomalies/patterns.  

\point{Defence 3: \FailSafe Interceptor Service: 
Fortune is on the attacker’s side thus far} The victim proceeded without leveraging FBR, being lured into sharing the seed phrase or signing a transaction of the attacker’s choice.  At this point all that remains is to submit the transaction so that its effect will be reflected as part of the next block of the public ledger.  The attacker may choose to submit to any participating network node to be queued up with other pending transactions in the public memory pool (the holding area used to prioritise, and order proposed transactions for the next block on the ledger).

At this stage, the \FailSafe Interceptor Service (FIS) is on standby.  While monitoring a low latency stream of ingress mempool transactions, FIS filters for transactions that\FailSafe users partake in.  Once detected the counterparty address is passed to FBR.  If a threat is detected, FIS attempts to make the attacker’s transaction revert. It should be noted that FIS intervention may also be triggered by transaction policy limits that are part of the user’s\FailSafe configuration (e.g., max value allowed to be transferred in a given time period, etc.).

To intercept, FIS submits a new transaction into the pool that transfers the assets in play to another address owned by the user (e.g., the cold wallet address). Most importantly this new transaction will have a slightly higher gas price than that being paid by the attacker, which results in being placed ahead of the attacker's transaction, in the execution order.  When it comes time to execute the attacker's transaction, it will revert and not become part of the ledger as the user will have an insufficient balance at that point.  The same principle applies to NFTs~(ERC-721).

To improve the odds, the attacker may choose to pay a transaction fee premium and bypass the global memory pool altogether through so-called ``private transactions.''  These are aggregated in bulk via  intermediaries and included by miners in the next mined block) on the blockchain.

To close down this avenue to fraudsters and apply the above techniques, we are exploring an ``exceptions list'' mechanism, where end-users will be able to add their own web3 address to this list.  Mempool service providers such as BloxRoute would then filter out private transaction requests where the source address is a member of this list.

Additionally, throughout the life-cycle of the transaction the user is alerted with real-time push message notification and further advice on possible next remediation steps.


\point{Summary}
Defences~1--3 provide the elements of defence-in-depth that the \FailSafe approach rests on. 



%% file: sections/03_architecture.tex
\begin{figure}[tbp]
\centering
\includegraphics[width=0.85\textwidth]{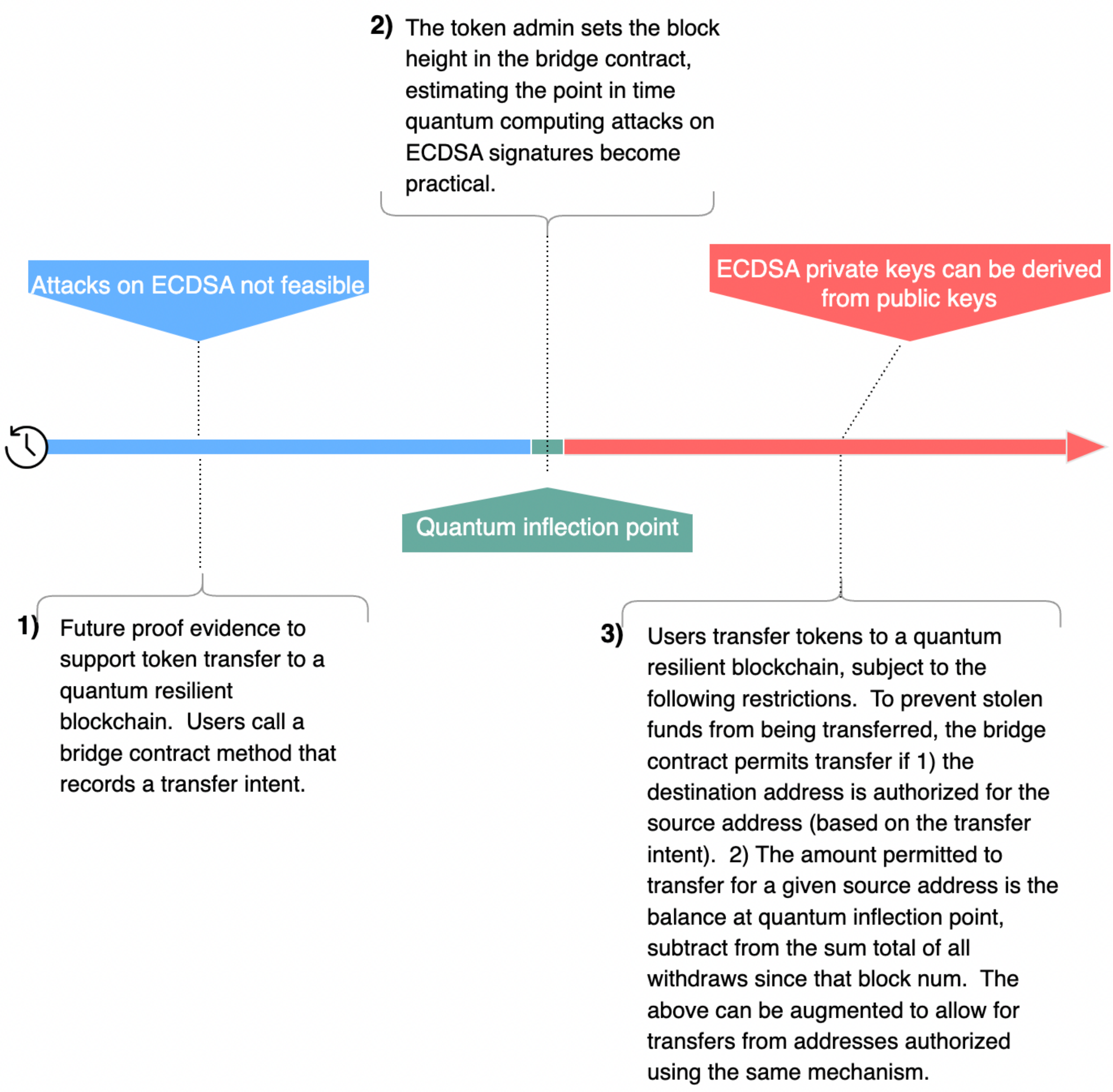}
\caption{Timeline and approach of \qMig.}
\label{fig:qmig}
\end{figure}

\begin{figure}[tb]
\centering
\includegraphics[width=1\textwidth]{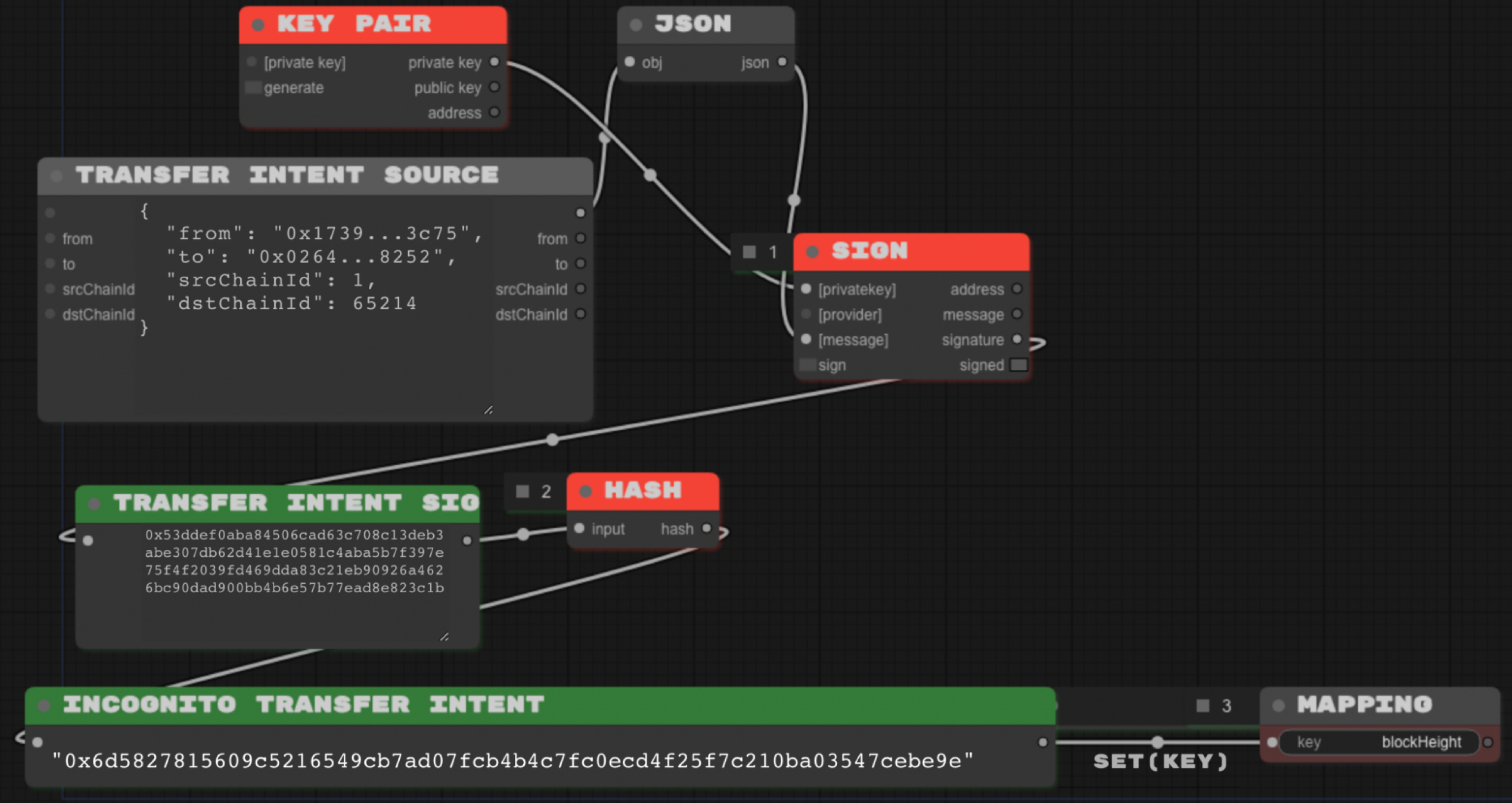}
\caption{Generating incognito transfer intent.}
\label{fig:generating-intent}
\end{figure}

\begin{figure}[tb]
\centering
\includegraphics[width=1\textwidth]{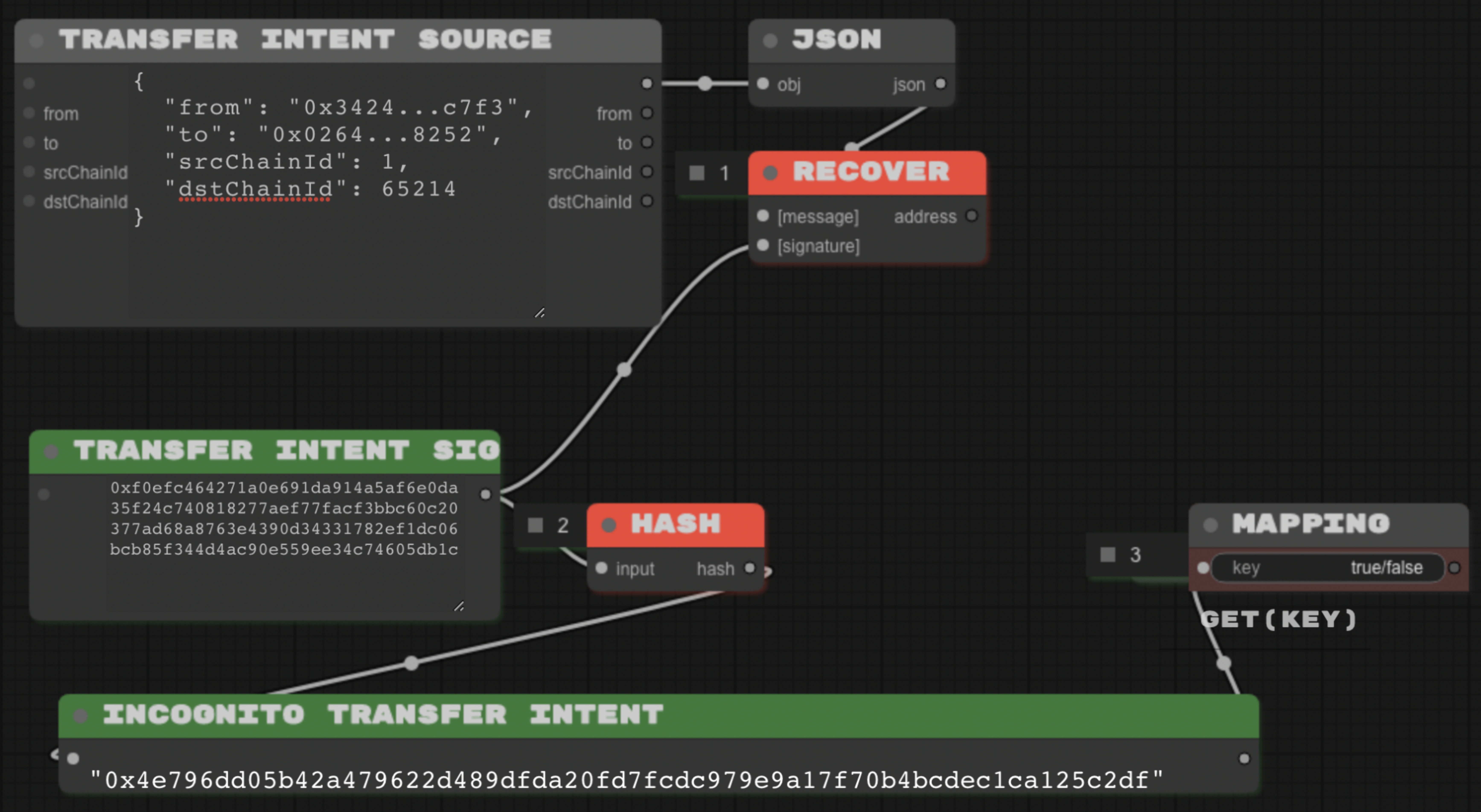}
\caption{Verifying incognito transfer intent.}
\label{fig:verifying-intent}
\end{figure}

\section{\FailSafe Architecture}
\label{sec:arch}

Figure~\ref{fig:arch} presents a high level view of the overall \FailSafe system architecture.  Starting at the top, as part of the enrollment process, a factory contract is used to deploy a dedicated user instance of the multi-signature \FailSafe contract.  As illustrated in the figure,  each operation supported by the contract may require one or more independently controlled keys for an authorization before a given operation is executed. For instance, in the example illustrated in the figure, the withdrawal operation (from the contract to the enrolled hot wallet key), requires three independent signatures.  

To call the \FailSafe contract’s \code{intercept} and \code{rebalance} methods, both the \FailSafe Interceptor Service and Assets Balancer Service use the dedicated enclave for signing, while access to private keys are subject to the above restrictions.  In the Multi-Sig \FailSafe contract these operations are configured to require a single signature since the assets are being shifted between addresses under the user control, based on a prior user consented to configuration.  In contrast, an administrative operation (e.g., updated ratio of hot/cold asset balances) or asset withdrawal, requires multiple independently controlled key signatures.  With \FailSafe, a user enrolls in the AWS Cognito Identity management system which supports multi-factor authentication~(MFA). 

The user's Cognito account ID is mapped to the user’s \FailSafe Account.  Similar to the intercept and rebalance private keys, the private key representing a given user's authentication factor has a corresponding key protected by Nitro Enclave/KMS (this is supported via Cognito identity pool/lambda function authentication flow).  To mitigate against worst-case insider attacks, the user has another authentication factor enrolled and protected by a secondary, independent cloud~(GCP), which offers similar key isolation and protection environment via Confidential Compute + KMS with cryptographic attestation proofs.

Next, let's consider how the keys for these operations are managed and protected.  As an example, let's consider the intercept key which, as described in the previous section, is required for moving funds from the user’s hot wallet address to the user's \FailSafe contract in the event that a pending fraudulent transaction is detected in the memory pool.  As illustrated in Figure~\ref{fig:arch}, in the cloud, at rest the intercept key is only only preserved in an encrypted form.  It is encrypted under the data encryption key that resides in the AWS KMS/Cloud HSM (hardware security module).  

The \FailSafe system leverages the AWS Nitro Enclave + KMS security architecture.  The request to decrypt the intercept key is fulfilled by KMS if~1) the request comes from the expected IAM identity;~2) the request is made from the target Nitro enclave. AWS Nitro Enclave is a hardened, isolated execution environment; cryptographic attestation is used to prove to KMS the enclave's identity and the code running in the enclave.  

\subsection{Forward Security in \FailSafe}
As part of the \FailSafe project, the Quantum Migration Tool was developed to address future platform level threats to the user's web3 assets. This section examines the threats stemming from quantum computing to EVM based blockchains. We then present the design of the quantum migration tool and its role in the overall architecture of the \FailSafe system.

\subsection{Quantum Threats to EVM-based Blockchains}
Shor's Algorithm~\cite{shor_polynomial-time_1997} makes it possible for a sufficiently powerful quantum computer to break the ECDSA algorithm. That is, starting from a transaction signed with an ECDSA private key, one can extract the public key and then derive the private key.  This is the ultimate game over condition, as the attacker can then transfer any balance associated with the external owned account~(EOA) at will.

In contrast, quantum computers pose no such (known) threat to hashing algorithms.  Grover’s algorithm~\cite{grover_fast_1996} (aka quantum search algorithm), reduces the search for collisions in Keccak-256 (Ethereum’s hash algorithm) from $2^256$ to $2^128,$ which is less efficient than some generic collision search algorithms.  (A quick peek ahead: this hashing resilience to quantum attacks will play a key role in our approach).

For the underlying cryptography, the National Institute of Standards and Technology~(NIST) initiated a standardisation effort for quantum resilient signature schemes, and is currently evaluating a number of candidate schemes.  All of these come with their own set of trade-offs, particularly when compared with key size, speed and re-use of the same key pair by the EVM family of blockchains~\cite{westerbaan_nists_2022}.

In terms of the threat timeline, (i.e, how long until quantum hardware is capable of breaking ECDSA), estimates vary between experts.  Many believe the threat is still in the distant future (e.g., Vitalik was famously quoted comparing quantum computing advances to going from hydrogen bombs to harnessing nuclear fusion).  

For a systematic approach, the Global Risk Institute conducts an annual survey on the threat timeline of leading subject matter experts.  According to its~2022 report, the “likelihood” estimates have been trending upwards from initial surveys.  Nearly~25\% of respondents estimated a~50\% chance for the threat to materialise within a~10-year time window in light of recent advances (i.e., Google’s Quantum Supremacy and IBM Quantum System One) and the nation state competition (aka ``quantum race'') with high levels of funding. The inevitable question, much like the plight of global warming, is not a question of ``if''~---~it's a question of ``when.''

\subsubsection{On ECDSA Key Re-use}
Networks where the common pattern is to reuse the same key pair across different transactions (like the EVM family of blockchains) face a greater risk, once quantum attacks become feasible. The attacker has a longer time window to derive the private key via the earlier record of transaction signatures. However, using new ECDSA key pairs per-transaction, may only offer some temporary relief; once quantum attacks become sufficiently fast, an attacker could derive the private key and front-run a targeted transaction. 

\subsubsection{Account Abstraction as a Path to Sunseting ECDSA on Ethereum?}
A future version of Ethereum is expected to support an account abstraction, a unified representation of an account (rather than the two types that exist today; a smart contract account, and an externally owned account~(EOA) with a corresponding ECDSA private key). The most recent account abstraction proposal under consideration is EIP-4337~\cite{erc4337}.  Among its features is a  representation of an account as a smart contract wallet with cryptographic agility for submitting requests (referred to as \code{UserOperations}) to the wallet.  \code{UserOperations} can be signed using a quantum safe signature scheme.  Under this proposal, after a network upgrade, the current user base with quantum vulnerable EOAs ``can individually upgrade their wallets to quantum-safe ones,'' as noted by Vitalik.

In the event of a quantum attack breakthrough, the user-dependent upgrade strategy might mean a large number of unconverted accounts.  Addresses with prior transaction history would be at highest risk. On Ethereum, by design, externally owned addresses are commonly re-used.  Addresses with a prior transaction history, the ECDSA public key can then be readily retrieved.

If the quantum attack breakthrough occurs while any significant portion of EOAs haven’t been upgraded yet, any subsequent transactions signed with ECDSA (including upgrade to quantum resilient wallet) would be suspect: is it the attacker or the key rightful owner performing the operation? By comparison, this dilemma is more severe than the Ethereum rollback debate after the~2016 DAO exploit~\cite{clasic_ethereum} (which resulted in a hard fork, and two chains going forward, Ethereum and Ethereum classic).

To address this problem, a path rooted in cryptographic based trust is needed even when the algorithm authorising the majority of today's transactions is compromised (i.e., in a scenario where users need to migrate their web3 assets to a forked version of the chain, where UserOperations only use quantum resilient algorithms). 

\subsection{Quantum Migration Tool (\qMig)}
\label{sec:qmig}
\subsubsection{Assumptions and Goals}

\begin{itemize}
\item The quantum inflection point (quantum attack breakthrough)  may occur when the large majority of web3 assets and transactions are done on blockchains reliant on ECDSA.
\item Web3 tokens (ERC-20, ERC-721, etc.) are only allowed to be bridged (e.g., via LayerZero) to a quantum-resilient network,  if and only if  these assets have not been hijacked via ECDSA compromise.  That is, this transfer must be based  on a cryptographic scheme that remains secure even after the above quantum inflection point occurs.
\item While users could reduce their own exposure to the attack by not reusing the same address across transactions, it is not a prerequisite for using \qMig. \qMig should prevent movement of stolen funds, while supporting EOAs with prior transaction history.
\end{itemize}

\subsubsection{The Workings of \qMig}
\point{Step 1}
The \qMig approach is illustrated with a timeline shown in Figure~\ref{fig:qmig}. The blue phase represents a time period when quantum attacks on ECDSA are still not feasible.  In this period, the \qMig contract enables users to construct and record a future intent to transfer tokens, in case the quantum inflection point occurs and ECDSA signature by itself can not be trusted.  The security of this intent will be rooted in cryptography that is not susceptible to quantum attacks.  Figure~\ref{fig:generating-intent} illustrates how this works.  To call \qMig.\code{registerTransferIntent()}, the client creates a transfer intent source structure that includes source EOA (\code{from} field), the chain ID for this EOA and its future destination counterpart.  So for example, the source address \address{0x1739…3c75}, may own USDC (ERC-20 token) on chain ID~1 (Ethereum mainnet).  An updated bridge contract (e.g., LayerZero), will be able to utilise this information, to verify that the destination address was authorised by \address{0x1739…3c75}, prior to the inflection point.

As shown in Figure~\ref{fig:generating-intent}, the client must sign the transfer intent source with the ECDSA private key that corresponds to the EOA source address in the structure (this linkage is verified at a later stage, if transfer is initiated in step~3). 

The signed output is fed into Keccak-256 hash function, the resulting digest is referred to as incognito transfer intent in the above figure.  This digest is stored in a hash table by \code{registerTransferIntent()} method along with the corresponding block height (e.g.,~\code{16316192}) , preserving a record on the blockchain of when the intent was registered. 
Please note:
\begin{itemize}
\item The hash of the signature rather than the signature itself is persisted so that a public key of EOA can not be extracted.  Moreover the wallet address used to sign the request for calling \code{registerTransferIntent()} is recommended to be different from the source EOA, to avoid leaving a record of the ECDSA public key.
\item The Keccak-256 function is assumed to be resilient to finding collisions based on quantum hardware.
\item After the inflection point, the recorded block height along with the hash over the signature will serve as proof that source EOA authorised the transfer to the target destination address and not the attacker’s address. The ledger record shows that this was captured, prior to the quantum breakthrough when the attacker could derive the ECDSA private key.
\end{itemize}

\point{Step 2}
ECDSA is compromised (the quantum inflection point of Figure~\ref{fig:generating-intent}). For every token contract prepared for this scenario, its token administrator sets the block height in the bridge contract, representing a point when quantum computing attacks on ECDSA are effective.  If this value is set, a new authorization policy takes effect for all further transfer of assets over the bridge to a quantum safe network.  It is described in step~3.  Note that the bridge contract's administrative functionality should be protected by a secondary, quantum-resilient signature (e.g., CRYSTALS-Dilithium~\cite{computer_security_division_announcing_2022}). 

\point{Step 3}
Represented as the red phase in Figure~\ref{fig:generating-intent}. In this phase, users can transfer tokens to a quantum-resilient blockchain, subject to the following restrictions, to avoid stolen funds from being transferred:

\begin{itemize}
    \item The destination address was authorised by the source address from step~1. That is, at some point prior to the inflection point, the user needs to call \code{registerTransferIntent()}.  The bridge contract can verify this by calling the \code{verifyTransferIntent()} method exposed by the \qMig contract.  To achieve this, the client will build and sign the transfer intent source (shown in figure~\ref{fig:verifying-intent}, and then will pass the  raw signature (transfer intent sig) in conjunction with the transfer request to the bridge contract.  The bridge contract calls \code{verifyTransferIntent()}, passing the source and destination info, the raw signature (transfer intent sig) as well as the blockchain height value (the inflection point set in step~2).  
    
    \code{verifyTransferIntent()} performs verification through a process shown in Figure 5.  It uses the public key from the transfer intent sig to cryptographically verify the signed data and confirm the signer’s address matches the source address in the transfer intent source structure, thus confirming that it was the source that authorised the destination address.  Next, is the look up to check that the \qMig contract has a record of the target intent.  Just like in step~1, the transfer intent sig is passed to Keccak-256, to compute the incognito transfer intent digest.  This value is then looked up in a hash table of all recorded transfer intents.  If found, the returned block height must be less than the block height passed in by the bridge contract:
\begin{verbatim}    
	require (intentBlockHeight  < inflectionPointBlockHeight, 
		''Intent to transfer registered after the quantum inflection point!'')
\end{verbatim}

\item The next key requirement to address is that this mechanism must prevent any potentially stolen assets (due to ECDSA compromise) from being bridged to the quantum-safe blockchain.  The source address in the transfer intent could have funds obtained illicitly from others via ECDSA compromise, sometime after quantum inflection point.  To address this, the amount permitted to transfer for a given source address is the balance at quantum inflection point (block height set in step~2), subtract from the sum total of all withdraws since that block number (up to the most recent block). The above can be augmented to allow for transfers from addresses that were explicitly authorised by the destination addresses to source address prior to the inflection point (using the same mechanism).
\end{itemize}

\subsection{Combining \FailSafe and \qMig}
The \qMig approach is a low friction means for users to prepare for accelerated breakthroughs in quantum hardware based attacks.  The user can continue to conduct business on today's networks, while setting up a path to migrate assets to a quantum safe network if needed. The intent to transfer can be implemented and recorded on a \qMig contract on today's chains.  It is also feasible to build out bridging support to an existing quantum ready network like QVL in the near term. The same infrastructure is then applicable to quantum safe versions of EVM networks once they become available.

Users enrolled into the \FailSafe system are afforded the quantum threat protections developed as part of the \qMig tool.  This is facilitated in two steps. As part of wallet address enrollment in the user's \FailSafe contract shown in Figure~\ref{fig:arch}, a separate call is made by the client to register an incognito transfer intent with the quantum migration proofs contract (as described in the previous section). Secondly, as part of the enrollment transaction, the user’s \FailSafe contract calls the \qMig contract, registering an intent to transfer funds from the \FailSafe contract to one or more user wallet addresses registered with the contract. 

After the quantum inflection point, during the bridging of assets to a quantum resilient network, the \qMig contract can then take into account any transfers from the contract back to the wallet. To prevent fraudulent transfers due to ECDSA cryptography compromise, recall that the amount permitted to transfer for a given source address is the balance at quantum inflection point, subtract from the sum total of all withdraws since that block number (up to the most recent block) and adjusted for any other transfer intents to the source address that were registered prior to the inflection point.

%% file: sections/04_related.tex
\section{Related Work}
\label{sec:related}

\subsection{Wallet Security}
Below we describe some of the recent academic work focusing on wallet security, although the reader is referred to industrial reports\footnote{\url{https://www.continuumloop.com/wp-content/uploads/2022/02/The-Current-and-Future-State-of-Digital-Wallets-v1.0-FINAL.pdf}},\footnote{\url{https://www.mdpi.com/2076-3417/12/21/11180/pdf}},\footnote{\url{https://go.sensortower.com/rs/351-RWH-315/images/state-of-crypto-apps-in-europe-2022.pdf}}.

Karantias~\cite{sok-wallets}  provides the first definition of a cryptocurrency wallet, which they model in a client-server paradigm. The authors categorize wallets based on whether they work for transparent or private cryptocurrencies, what trust assumptions they require, their performance and their communication overhead. For each type of wallet the authors provide a description of its client and server protocols. They explore superlight wallets and describe their difference to superlight clients that have appeared in recent literature.The paper demonstrates how new wallet protocols can be produced by combining concepts from existing protocols. Finally, the paper evaluates the performance and security characteristics of all wallet protocols and compare them.

By analyzing source code, bytecode, and execution traces, di Angelo~et al.~\cite{wallet-contracts} derive usage scenarios and patterns. The authors discuss methods for identifying wallet contracts in a semi-automatic manner by looking at the deployed bytecodes and the on-chain interaction patterns. The authors extract blueprints for wallets and compile a ground truth. Furthermore, the paper differentiates characteristics of wallets in use, and group them into several types. The paper provides numbers and temporal perspectives regarding the creation and use of wallets. For the~40 identified wallet blueprints, the authors compile detailed profiles. The authors analyze the data of the Ethereum main chain up to block~11,500,000, mined on~December~22,~2020.


\subsection{Usable Security for Blockchain Users}
Issues related to usable security in the cryprocurrency space are not particularly well understood, although some studies are starting to appear. 

Specifically, Froehlich~et al~\cite{Froehlich_2022} provide a literature survey focusing on usable privacy and security for cryptocurrency, although a lot of the focus has been around user's perceptions of privacy in light of anonymization that can be accomplished through address clustering or more subtle attacks around mixers and and privacy-friendly chains like Monero. We  expect that new reports will start to appear in the near future, focusing on end-user interactions, for instance~\cite{preferring-stablecoin}.

Mangipudi~et al.~\cite{uncovering-impact} look at mental models around wallet usage. This work presents a data-driven investigation into the perceptions of cryptocurrency users towards multi-device wallets today,  using a  survey  of~255 crypto-wallet  users. Their results revealed two significant groups of participants: Newbies and Non-newbies. These two groups statistically significantly differ in their usage of crypto-wallets. However, both of these groups were concerned with the possibility of their keys getting compromised and yet are unfamiliar with the guarantees offered by multi-device wallets. After educating the participants about the more secure multi-device wallets, around~70\% of the participants preferred them; However, almost one-third of participants were still not comfortable using them. The qualitative analysis revealed a gap between the actual security guarantees and mental models for these participants: they were afraid that using multi-device wallets will result in losing control over keys (and in effect funds) due to the distribution of key shares. Moreover, considerations about the threat model further affected their preferences, signifying a need for contextualizing default settings. 

Min~et al.~\cite{min2022portrait} attempt to portray DApp users through large-scale Ethereum data, seeking to present an understanding of the data aspects of users in the blockchain scenario beyond surveys and interviews.They built a series of datasets labeled with DApp name and extracted information about each address to enrich the data dimension. We then visualized and analyzed the user profile dataset from several aspects  and explored methodologies to divide user groups. The authors propose a way to use the number of interactions by the categories of DApp as cluster input and classify the addresses with SOM network. After that, they separated active addresses into four groups according to the purpose and frequency of use and discuss the differences between them and their sensitivity to the market. In addition, the paper gives examples of how to use transactional  data. The authors combine the results of their analysis with previous research studies to summarize the profile of DApp users and examine user  motivations, values, and the current DApp market.   

Ghesmati~et al.~\cite{user-perceived-privacy} focus on how well end-users understand privacy guarantees in the blockchain space. They conducted a study on user perception and preference on Bitcoin privacy, investigating different add-on privacy techniques in Bitcoin as well as their implementation in practice. The authors showed the difference between users' preferences and the implementation of privacy techniques in practice. Most users preferred privacy coins rather than add-on techniques in Bitcoin. The results show that participants are more likely to accept delays rather than extra fees to achieve anonymity in Bitcoin. The participants also preferred indistinguishable privacy techniques rather than being flagged by monitoring tools. Therefore, important questions are raised as current privacy wallets offer CoinJoin transactions with equal-sized output that are distinguishable in the blockchain. They show that users who prefer better privacy are not likely to use Bitcoin, and they favor embedding built-in privacy features in Bitcoin.

Korir~et al.~\cite{decentralized-identity} focus on the prevalence of decentralized identify approaches. There is a growing expectation that political and technical initiatives towards digital identity will gather pace in the foreseeable future. However, user perspectives have not been a
driving force in shaping those ongoing initiatives. The findings of this study point to the dominance of paper/card-based identity methods for online identity verification and a large gap between identity verification today and what it might be in the foreseeable future. The results suggest that technical narratives might not be a compelling driving force for future
uptake and that, as previous work in identity management has
highlighted, the user proposition should receive further thought. What seems most salient to drive adoption is the existence of supporting (infra)structures, the appeal of the list of available verifiers, and the low complexity of using a new identity wallet tool. 

Mai~et al.~\cite{mental-models} explores user perceptions and misconceptions of cryptocurrency users ($N$~=~29) enriched with drawing and card assignment tasks. Although the study focused on Bitcoin and Ethereum, the findings can be further useful for improving the security and privacy of a large body of (existing or future) altcoins which also build on the blockchain technology. The paper points out that flaws and inconsistencies in user mental models of cryptocurrency systems expose users to security and privacy risks when using current cryptocurrency tools. These risks include money loss, fraud, or deanonymization. Most importantly, the paper revealed major misconceptions related to the functionality and management of cryptographic keys which are not compensated by the cryptocurrency tools. The findings explain why cryptocurrency users fail to manage their private keys securely and, as a result, frequently fall victim to money loss and fraud. Furthermore, users think that the blockchain is encrypted or oblivious, which prevents them from taking measures to safeguard their privacy. Another interesting result was that many participants were not aware of the fact that the amount of mining fees can be actively selected to influence the transaction speed. They proposed several concrete enhancements to state-of-theart cryptocurrency tools (e.g., wallets or exchanges) with the purpose of protecting users with misconceptions from security and privacy threats. Among others, the paper suggests suggest automation of key generation, management, and back-up as much as possible. With this work, the authors create a foundation for improving the usability of state-of-the-art cryptocurrency management tools to prevent security and privacy breaches. 


A recent report by Wang~et.~al.\cite{WangFWZ0Y22} highlights some of the common usage practices around Web3 wallets and provides compelling motivation for \FailSafe.
In this paper, the authors present the first in-depth study of quantifying the risk of unlimited approval of ERC20 tokens on Ethereum. The study proposes a fully-automatic approach to detect the approval transactions, and reveals the high prevalence~(60\%) of unlimited approval in the ecosystem. The authors conduct an investigation to reveal the security issues involved in interacting with 31 UIs (22 DApps and 9 wallets) to send approval transactions. 

The result shows that only a few UIs provide explanatory understandings~(10\%) and flexibility~(16\%) for users to mitigate the risk of unlimited approval. Furthermore, the paper performs a user behavior analysis to characterize five modes of user behaviors and formalize the good practice to use approved tokens. The result reveals that users~(0.2\% of user behaviors) barely follow the good practice towards mitigating the risks of unlimited approval. Finally, the paper discusses two existing solutions attempting to address the trade-off between convenience and security of unlimited approval, and provide possible suggestions.


\subsection{Quantum-resistant Blockchains}
Post-quantum cryptography (required for quantum resilient blockchains), is being standardized by the National Institute of Standards and Technology~(NIST).  The NIST standardization effort for quantum resilient signature schemes, and is currently evaluating a number of candidate schemes (CRYSTALS-Dilithium, FALCON \& SPHINCS).  All of these come with their own set of trade-offs, particularly when compared with key size, speed, and re-use of the same key pair by the EVM family of blockchains~\cite{westerbaan_nists_2022}.

A future version of Ethereum is expected to support an account abstraction, a unified representation of an account (rather than the two types that exist today; a smart contract account, and an externally owned account~(EOA) with a corresponding ECDSA private key). The account abstraction model also aims to provide cryptographic agility, which would allow for post quantum safe signature algorithms.

\begin{itemize}
\item The account abstraction architecture undergone several different iterations. The first proposal dates back to~2016 with EIP-86~\cite{erc86}.  It was primarily focused on converting all accounts into contracts, with a more flexible security model than the current version of Ethereum (i.e., support for multisig and a framework to specify alternative signature algorithms, beyond ECDSA).

\item The next version of account abstraction was proposed in EIP-2938~\cite{erc2938}.  Which was in part motivated by gas efficiency and introduced new transaction types and corresponding new opcodes. Next, EIP 3074~\cite{erc3074} proposed a mechanism by which an externally owned account can delegate control to a smart contract, enabling delegation of fees for block chain transactions.

\item The most recent account abstraction proposal under consideration is EIP-4337~\cite{erc4337}.  Among its features is a  representation of an account as a smart contract wallet with cryptographic agility for submitting requests (referred to as \code{UserOperations}) to the wallet.  \code{UserOperations} can be signed using a quantum safe signature scheme.  Under this proposal, after a network upgrade, the current user base with quantum vulnerable EOAs ``can individually upgrade their wallets to quantum-safe ones,'' as noted by Vitalik.
\end{itemize}
In the event of a quantum attack breakthrough, the user-dependent upgrade strategy might mean a large number of unconverted accounts.  Addresses with prior transaction history would be at highest risk. On Ethereum, by design, externally owned addresses are commonly re-used.  Addresses with a prior transaction history, the ECDSA public key can then be readily retrieved.

If the quantum attack breakthrough occurs while any significant portion of EOAs haven’t been upgraded yet, any subsequent transactions signed with ECDSA (including upgrade to quantum resilient wallet) would be suspect: is it the attacker or the key rightful owner performing the operation? By comparison, this dilemma is more severe than the Ethereum rollback debate after the~2016 DAO exploit~\cite{clasic_ethereum} (which resulted in a hard fork, and two chains going forward, Ethereum and Ethereum classic).

A notable effort to build an entirely new quantum resilient blockchain is Quantum Resistant Ledger (QRL)~\cite{qrl}, which uses XMSS~\cite{huelsing_xmss:_2018} signatures.  An affiliated project, EnQlave, is a wallet  implemented as an Ethereum smart contract that uses a secondary signature~(XMSS), resilient to quantum attacks; the scheme relies on a Merkle tree of cryptographic hashes. The idea is to transfer your tokens into this quantum resilient vault as a means to mitigate future unexpected events (i.e., a successful quantum attack).  

While this mechanism is useful to protect individual EOA, it does not deal with the resulting fallout of such an attack: a loss of confidence in a system rooted in cryptographic trust. Any further transaction signed with an ECDSA key would be suspect.  Business as usual on a network in this state would likely cease.  In case of such an event, there is a need for a mechanism that can securely move assets that belong to their rightful owner (rather than the attacker) to a quantum-resilient network.

%% file: sections/05_conclusions.tex
\section{Conclusions}
\FailSafe provides a  set of safety and protection mechanisms that are built for a user base that does not necessarily follow best security practices at all times.  \FailSafe avoids over reliance on any single defence mechanism, if one is bypassed, next one is inline to help minimise losses.  This approach spans across the entire lifecycle of a Web3 transaction, from de-risking Web 3 asset positions (via auto-rebalancing asset to cold storage) to intercepting the attacker’s transaction via the blockchain mempool, if all other defences failed. 

Similarly, with \FailSafe’s forward-security, the risk to the user’s Web~3 assets are reduced even if the underlying cryptographic based trust is compromised.